**Long-Range Nanoelectromechanical Coupling at the LaAlO$_3$/SrTiO$_3$ Interface**

Aditi Nethwewala[1,2], Kitae Eom[3], Muqing Yu[1,2], Ranjani Ramachandran[1,2], Chang-Beom Eom[3], Patrick Irvin[1,2], Jeremy Levy[1,2*]

[1]Department of Physics and Astronomy, University of Pittsburgh, Pittsburgh, PA 15260, USA.

[2]Pittsburgh Quantum Institute, Pittsburgh, PA, 15260 USA.

[3]Department of Materials Science and Engineering, University of Wisconsin-Madison, Madison, WI 53706, USA.

[*]To whom correspondence should be addressed. E-mail: jlevy@pitt.edu

**Abstract:**

The LaAlO$_3$/SrTiO$_3$ interface hosts a plethora of gate-tunable electronic phases. Gating of LaAlO$_3$/SrTiO$_3$ interfaces are usually assumed to occur electrostatically. However, increasing evidence suggests that non-local interactions can influence and, in some cases, dominate the coupling between applied gate voltages and electronic properties. Here, we sketch quasi-1D ballistic electron waveguides at the LaAlO$_3$/SrTiO$_3$ interface as a probe to understand how gate tunability varies as a function of spatial separation. Gate tunability measurements reveal the scaling law to be at odds with the pure electrostatic coupling observed in traditional semiconductor systems. The non-Coulombic gating at the interface is attributed to the existence of a long-range nanoelectromechanical coupling between the gate and electron waveguide, mediated by the ferroelastic domains in SrTiO$_3$. The long-range interactions at the LaAlO$_3$/SrTiO$_3$ interface add unexpected richness and complexity to this correlated electron system.

The ability to electrostatically tune semiconductor heterostructures and nanostructures is central to a host of scientific phenomena like the integer and fractional quantum Hall effect, and device applications such as field-effect transistors. Over the decades since the invention of the semiconductor heterostructure [1], new materials have been created and their properties explored through electrostatic gating. Complex oxide heterostructures with unit-cell control over composition were developed in the early 2000s [2], and the $LaAlO_3/SrTiO_3$ (LAO/STO) heterostructure [3] emerged as a system capable of many striking phenomena including metal-insulator transition [4], superconductivity [5, 6], magnetism [7, 8], and spin-orbit coupling [9], all of which are gate-tunable. Over the years, there have been intense efforts to understand the effect of voltage gating on the electronic properties at the LAO/STO interface [10]. A variety of gating techniques have been explored, such as back gating [11], top gating [12], proximal side gates [4], and ionic liquid gates [13, 14]. However, in most cases gating has been treated as a purely electrostatic phenomenon [15].

STO, a complex oxide, possesses a wealth of physical properties not commonly found in most semiconductors. STO undergoes a cubic-to-tetragonal phase transition at $T \sim 105$ K leading to the formation of ferroelastic domains along the crystallographic directions X, Y, and Z [16, 17]. Even though these ferroelastic domains are not intrinsically polar, there is strong evidence of coupling between ferroelastic domain structure and itinerant carriers at the LAO/STO interface, including enhanced conductivity along the ferroelastic domain walls [18, 19], giant piezoelectricity [16], and a non-universal current flow around the metal insulator transition [20]. Scanning single-electron transistor (SET) measurements at the LAO/STO interface show that ferroelastic domain walls can move over large distances under applied back gate voltages [16]. In reports by Frenkel et al. [21], localized pressure applied to polar domain walls is also shown to modulate the current distribution.

Mesoscopic devices have helped to reveal the connection between ferroelastic domains and the electronic properties at the LAO/STO interface. Experiments with sketched LAO/STO devices suggest a fundamental 1D nature of pairing and superconductivity, with the ferroelastic domain boundaries playing a key functional role [22]. Frictional drag experiments in LAO/STO nanowire system revealed long-range non-Coulombic electron-electron interactions [23] which differ from traditional semiconducting systems, and whose origin has been linked to ferroelastic domain structure [24-26]. Quasi-1D cross-shaped ballistic electron waveguides or "nanocrosses", sketched at the LAO/STO interface, show an inhomogeneous electronic structure, with characteristic features attributed to ferroelastic domain structures that are pinned to the nanocross geometry [27]. These nanocross devices also suggest an interplay between ferroelastic domains, anomalous Hall effect and electronic nematicity in LAO/STO [28].

Here, we seek to quantify long-range gating of nanostructures at the LAO/STO interface. We use quasi-1D ballistic electron waveguides as a probe of chemical potential shifts because of their characteristic transport "fingerprint", which enables quantitative information about changes in chemical potential that are induced by a variety of in-plane gates whose distance from the waveguide varies by more than an order of magnitude.

LAO/STO heterostructures are created by depositing a thin film of LAO (3.4-unit cell) on $TiO_2$-terminated STO by pulsed laser deposition, using growth conditions described elsewhere [29]. Measurements are performed with three different devices labeled A, B, and C. Electrical contact is made to the interface by depositing Ti/Au (4 nm/25 nm) electrodes surrounding a given "canvas". Eight interface contacts (labelled 1-8 in Figure *1*(a)) surround a 15 μm x 20 μm region on the canvas where devices are "sketched" with a voltage-biased conductive atomic force microscope (c-AFM) tip [30]. Additionally, two 11 μm wide interface contacts, "in-plane gates" (labelled 9 and 10 in Figure *1* (a)), are deposited 40 μm apart to gate the sketched devices near the center of the canvas. We also deposit two additional interface contacts (labelled 11-14 in Figure *1* (a)) on either side of the in-plane gates which are kept floating unless mentioned otherwise.

A ballistic electron waveguide [31] is sketched using c-AFM lithography. Conducting paths are created by applying a positive bias to the c-AFM tip, locally protonating the LAO surface, and attracting itinerant electrons to the interface. The insulating state is restored by applying negative voltages to the tip, which locally deprotonates the surface. The ballistic electron waveguide (depicted schematically in Figure *1* (b)) consists of a main channel of length $L_c = 100$ nm surrounded by two highly transparent tunnel barriers (width $L_b$ ~ 30 nm). The tunnel barriers decouple the main channel from the two terminal leads, allowing the electron density of the waveguide to be tuned by proximal side gates set to voltages $V_{sg1}$ or $V_{sg2}$ at a distance of $L_s$ ~ 2 μm, "up gate" voltage $V_{ug}$ at a distance of $L_u$ ~ 17 μm or "down gate" voltage $V_{dg}$ at a distance of $L_d$ ~ 23 μm [31]. Current $I$ is sourced between the source ($I_+$) and drain ($I_-$) and $V = V_{L+} - V_{L-}$ represents the four-terminal longitudinal voltage. All measurements are performed as a function of the applied gate voltage and an applied out-of-plane magnetic field, $\vec{B} = B\hat{z}$ at or near the base temperature of the dilution refrigerator, $T \sim 50$ mK.

Transport through quasi-1D ballistic electron waveguides is described by Landauer's formula $G = \frac{e^2}{h}\sum T_i(\mu)$ where each energy subband contributes one quanta of conductance $e^2/h$ with transmission probability $\sum T_i(\mu)$, $\mu$ being the chemical potential of the electron waveguide [31, 32]. The derivative of conductance $G$ with respect to $\mu$, also known as the "transconductance", reveals the subband structure of the waveguide. The number of energetically available subbands is controlled by the chemical potential which in turn is tuned by an applied gate voltage. The conversion factor that relates changes in gate voltage to changes in chemical potential known as the lever-arm ratio [33] can be determined by analyzing the nonlinear current-voltage relation of a device as a function of applied gate voltage.

The zero-bias longitudinal conductance $G = dI/dV$ of an electron waveguide primarily depends on the chemical potential $\mu$ and the applied magnetic field $\vec{B}$. In the given case, the chemical potential $\mu$ of the device can be tuned by the applied gate voltage, $\Delta\mu = \alpha_x \Delta V_{xg}$, where $V_{xg} = V_{sg1}, V_{sg2}, V_{ug}, V_{dg}$, and $\alpha_x$ is the measured lever arm for gate $x$. While the lever arm coefficients $\alpha_x$ may vary, the precise positioning of the gates is known to have a negligible impact on the observed subband structure of ballistic LAO/STO electron waveguides [31]. Figure 2 (e-h) shows the zero-bias longitudinal conductance, $G$ for Device A for magnetic fields ranging between $B = 0$ T and $B = 18$ T as a function of $V_{sg1}$, $V_{sg2}$, $V_{ug}$, and $V_{dg}$ respectively. Clear conductance quantization steps at $G = 2, 3$, and $4e^2/h$ is observed for all gate voltages applied. For the two proximal side gates at a distance of $L_s \sim 2$ µm, the electron waveguide in tuned from $G = 0 - 6e^2/h$ for $V_{sg1}$ and $V_{sg2} = 0 - 0.35$ V. Similar tuning of conductance is obtained between $0 - 4$ V for the up and down gates at $\sim 8 - 10$ times the distance of the proximal side gates.

Next, we calculate the lever arm ratio $\alpha_x = \frac{\Delta V_{4T}}{\Delta V_{xg}}$, with respect to the four gate voltages $V_{xg} \in \{V_{sg1}, V_{sg2}, V_{ug}, V_{dg}\}$. Figure 3(e-h) shows finite-bias spectroscopy performed through I-V measurements on Device A at $B = 18$ T as a function of the four gates, respectively. For each case, the horizontal red arrow marks the transition from one subband to another due to the applied bias $\Delta V_{4T}$. The energy gain induced by $V_{4T}$ is equal to the subband spacing marked by the vertical red arrow, $\alpha\Delta V_{xg}$ at zero bias, namely $e\Delta V_{4T} = \alpha\Delta V_{xg}$. Figure S1 and Figure S2 represent the $I$-$V$ measurements performed on Device B and Device C, respectively.

Table *1* summarizes the measured lever arm $\alpha_{sg1}$, $\alpha_{sg2}$, $\alpha_{ug}$ and $\alpha_{dg}$ for $V_{sg1}, V_{sg2}, V_{ug}$, and $V_{dg}$ for Devices A, B and C (color coded) respectively. The magnitude of $\alpha_{ug}$ and $\alpha_{dg}$, is always smaller than the corresponding magnitude of $\alpha_{sg1}$ and/or $\alpha_{sg2}$. The magnitude of $\alpha_{sg1}$, $\alpha_{sg2}$, $\alpha_{ug}$ and $\alpha_{dg}$ also significantly varies among the three devices and in the same device itself (See "$\alpha$" and "$\alpha * d$" headings in Table *1*). The spatial distances $L_s$ ~ 2 µm, $L_u$ ~ 17 µm and $L_d$ ~ 23 µm are kept constants for all devices. The variation among the devices is further highlighted by considering the ratio of the lever arms. Figure *4* shows the ratio of measured lever arm ratios for proximal side gates $\alpha_{sg1}$ and $\alpha_{sg2}$, versus the up gate, $\alpha_{ug}$ (left hand side) and down gate, $\alpha_{dg}$ (right hand side) for Devices A, B, and C. The ratio of $\alpha_{sg1}$, and $\alpha_{sg2}$, versus $\alpha_{ug}$ changes by a factor of three from $6 - 17.5$ among the three devices. Similarly, $\alpha_{dg}$ shows close to a two-fold increase from $12 - 21$.

We now turn our attention to the subband structure of the electron waveguide. The electronic structure is revealed by plotting the transconductance $dG/(dV_{xg})$ or $dG/d\mu$ as an intensity map as a function of $B$, and $V_{xg}$ or $\mu$. Peaks in the transconductance mark the chemical potential at which new subbands contribute to transport and are separated by regions where the conductance is constant ($dG/d\mu \to 0$) and ideally quantized at integer multiples of $e^2/h$. Figure *5* (e-h) shows the transconductance spectra $dG/(dV_{xg})$ as a function of $B$, and $V_{xg}$. The up and down gates which are at a distance of $L_u$ ~ 17 µm and $L_d$ ~ 23 µm respectively, reproduce similar subband structure as the proximal side gate $V_{sg1}$ and $V_{sg2}$ located at a distance of $L_s$ ~ 2 µm, with ~10 times greater voltage applied. However, $dG/d\mu$ plotted as a function of the chemical potential $\mu$, obtained from $\mu = \alpha V_{xg}$, shows similar order of chemical potential for all the applied gate voltages (see Figure *5* (i-l)). Thus, the chemical potential at which subband becomes available is comparable for all four cases, irrespective of the gate voltage applied.

To further highlight the observed variation in the magnitude and ratio of $\alpha_x$'s we consider the case of pure electrostatic gating. According to the constant interaction model the lever arm ratio is defined as [33, 34]:

$$\alpha_x = \frac{C_{xg}}{C_\Sigma},$$

where $C_{xg}$ is the geometrical gate capacitance and $C_\Sigma$ is the self-capacitance of the device. Self-capacitance can be assumed to be a constant for a given device. Hence, the lever arm ratio is solely determined by the geometrical capacitance, $C_{xg}$. Thus, e.g., the ratio of the lever arm of the proximal side gate and the up gate can be given by:

$$\frac{\alpha_{sg1}}{\alpha_{ug}} = \frac{C_{sg1}}{C_{ug}}$$

where $C_{sg1}$ and $C_{ug}$ are the geometrical capacitance of the side gate and the up gate, respectively. The given ratio can be used to determine the scaling law for electrostatic gating. For a homogeneous semiconductor system [35] with the same device geometry and parameters as considered here, the ratio $\alpha_{sg1}/\alpha_{ug}$ will be a constant value (See supplementary information for details). However, as shown in Figure *4*, $\alpha_{sgx}/\alpha_{ug}$ varies between 6 and 17.5 for the same set of devices sketched at the LAO/STO interface. Thus, gate tunability measurements on ballistic electron waveguides created at the LAO/STO interface reveals that the scaling of the lever arm with gate distance are at odds with a purely electrostatic interaction.

A possible scenario to explain the observed variation is to consider the influence of non-electrostatic contributions to the gating mechanism [36]. As mentioned earlier, the ferroelastic domains in STO are polar [21, 37, 38] and mobile [16] with strong coupling to applied gate voltages [16] and strain [39]. The applied gate voltages can further polarize and/or displace nearby ferroelastic domains, thereby introducing a nanoelectromechanical coupling between the gate and electron waveguide. Piezoelectric force microscopy measurements on LAO/STO nanostructures and transport measurements on LAO/STO nanocross devices [27] revealed that conducting nanowires in LAO/STO coincide with Z ferroelastic domains and are surrounded by insulating regions with X and Y ferroelastic domains. The surface potential of the Z domain differs from the X or Y domains by approximately 1 mV [16]. Thus, gating can produce both polarization and movement of X and Y domains, which can couple to the nanodomain structure of conductive nanowires in an electromechanical fashion. The strength of this coupling is not simply related to the physical proximity of the gate to the nanodevice but is instead governed by the three-dimensional ferroelastic domain structure, which is long-range and scales differently with what one would expect based on electrostatics alone [40]. Nanoelectromechanical gating due to ferroelastic domains at the LAO/STO interface deviates from the pure electrostatic behavior observed in semiconductor systems. The existence of long-range nanoelectromechanical coupling in LAO/STO suggests new types of coupling through, surface acoustic waves [41] and uniaxial strain [10]. The ability to control the three-dimensional ferroelastic domain structure will help reduce uncontrolled variations between devices, and lead to more precise understanding of the underlying mechanism of attraction that leads to electron pairing and superconductivity in STO-based heterostructures and nanostructures.

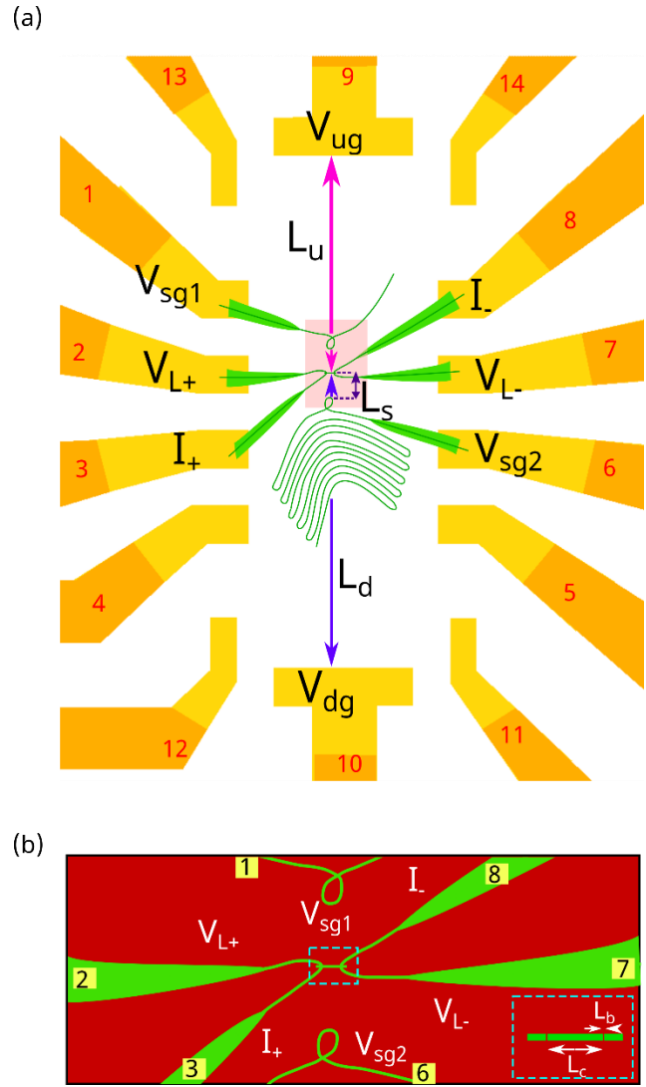

Figure 1: (a) Schematic showing gate and device geometry. The up gate is at a distance of $L_u \sim 17\ \mu m$, down gate is at a distance of $L_d \sim 23\ \mu m$ whereas the proximal side gates are at a distance of $L_s \sim 2\ \mu m$ from the main device respectively. Pink region shows the main device zoomed in panel (b), (b) Layout of the LAO/STO electron waveguide device. The ballistic nanowire consists of the main channel of length $L_c \sim 100\ nm$, with each end connected to two nanowire leads. Tunnel barriers of width $L_b \sim 30\ nm$ isolate the main channel from the two terminal leads, allowing the chemical potential to be tuned by the two available proximal side gates with voltage $V_{sg1}$ and $V_{sg2}$. Longitudinal voltage probes ($V_{L\pm}$) enable four-terminal conductance to be measured for current $I$ sourced between the source ($I_+$) and drain ($I_-$).

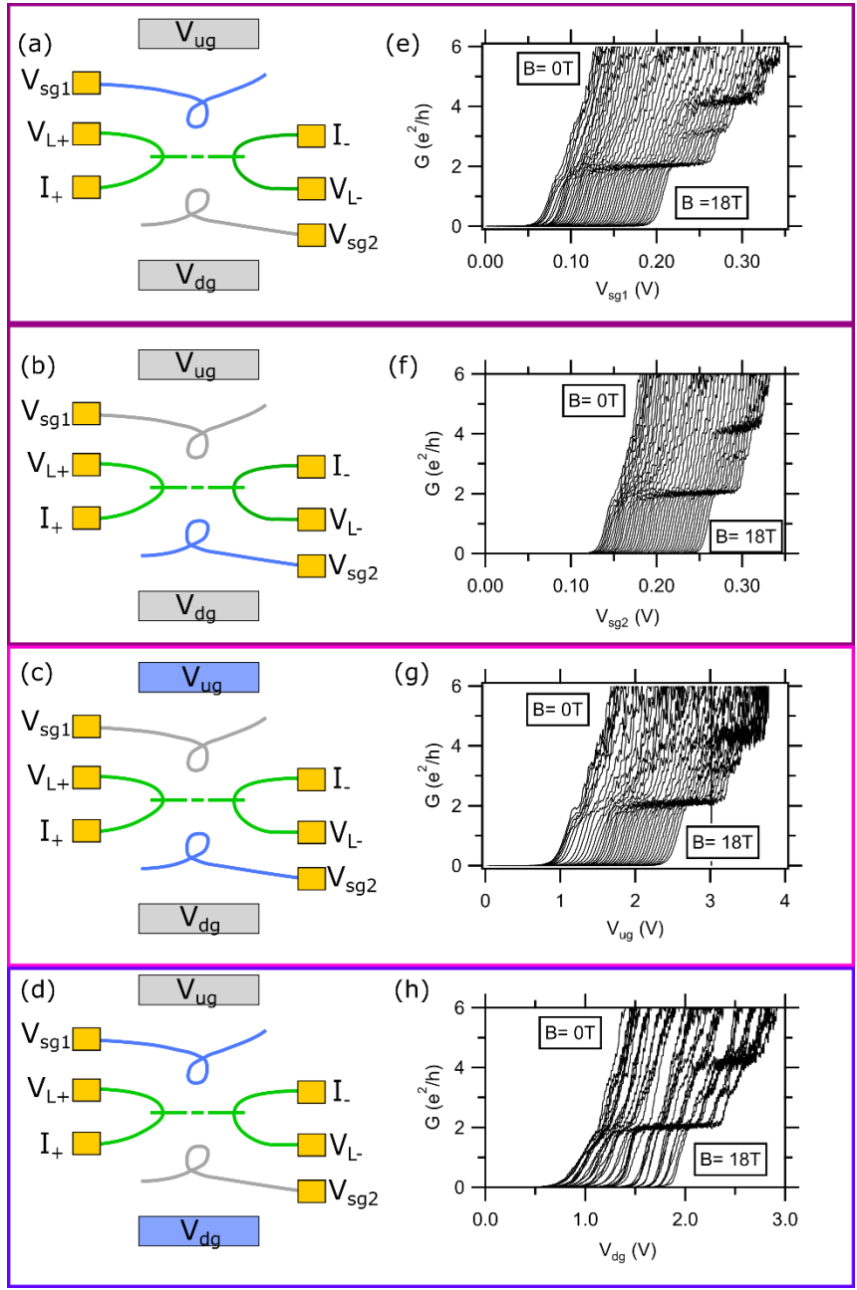

Figure 2: Longitudinal magnetoconductance measurements across device A for the four gate configurations (a-d) Green region shows the main device, blue region shows the gate voltages applied and grayed regions denotes that the gate is floating. In cases when the up and down gate voltages are being swept, the proximal side gate is maintained at a constant voltage for the whole sweep (b-h) Zero-bias longitudinal conductance, $G$, as a function of sidegate voltage, $V_{xg}$ and magnetic field $B$ in the range $0-18\,T$ for the four gate configurations, respectively. Data is shifted along x axis for clarity.

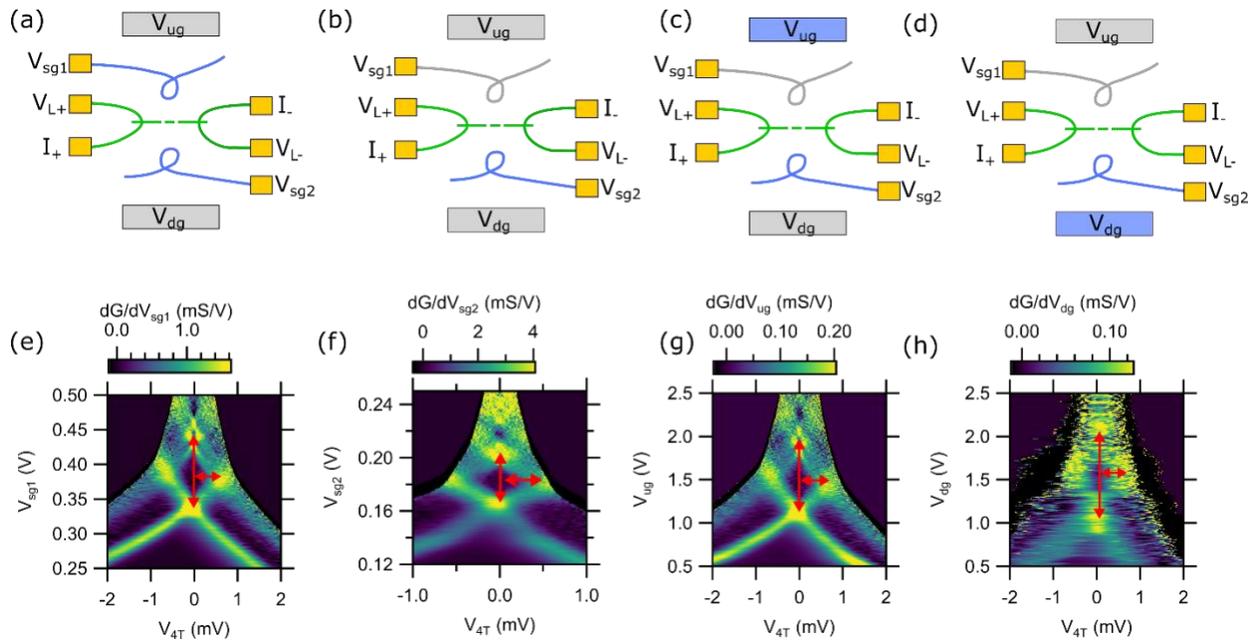

Figure 3: Finite Bias Spectroscopy on Device A (a-d) Schematic showing the device geometry and gate voltages applied. Green region shows the main device, blue region shows the gate voltages applied and grayed regions denotes that the gate is floating, (e-f) IV measurements across the different gate configurations denoted in panels (a-d). Red arrows denote the parameters used to calculate the lever arm for each configuration.

Table 1: Lever arm ratio with respect to proximal side gates, up gate and down gate for Devices A, B, and C

| Gate Type | Device | $d$ (μm) | $\Delta V_{4T}$ (V) | $\Delta V_{xg}$ (V) | $\alpha$ | $\alpha*d$ (nm) |
|---|---|---|---|---|---|---|
| Local side gate 1 ($V_{sg1}$) | A | 2 | 7.40E-4 | 8.90E-2 | 8.31E-3 | 17 |
|  | B | 2 | 5.30E-4 | 1.24E-1 | 4.27E-3 | 81 |
|  | C | 2 | 9.66E-4 | 3.56E-2 | 2.71E-2 | 54 |
| Local side gate 2 ($V_{sg2}$) | A | 2 | 5.33E-4 | 3.80E-2 | 1.40E-2 | 28 |
| "Up" gate ($V_{ug}$) | A | 17 | 6.64E-4 | 8.30E-1 | 8.00E-4 | 13 |
|  | B | 17 | 4.80E-4 | 6.80E-1 | 7.05E-4 | 12 |
|  | C | 17 | 7.27E-4 | 4.35E-1 | 1.67E-3 | 28 |
| "Down" gate ($V_{dg}$) | A | 23 | 5.86E-4 | 8.80E-1 | 6.66E-4 | 15 |

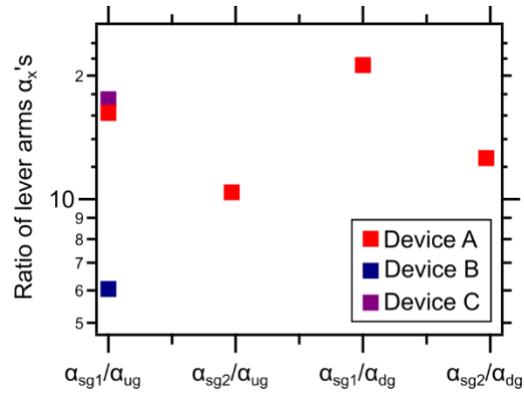

Figure 4: Variation in the ratio of the measured lever arms $\alpha_{sg1}, \alpha_{sg2}, \alpha_{ug}$, and $\alpha_{dg}$ for Devices A, B, and C on a logarithmic scale.

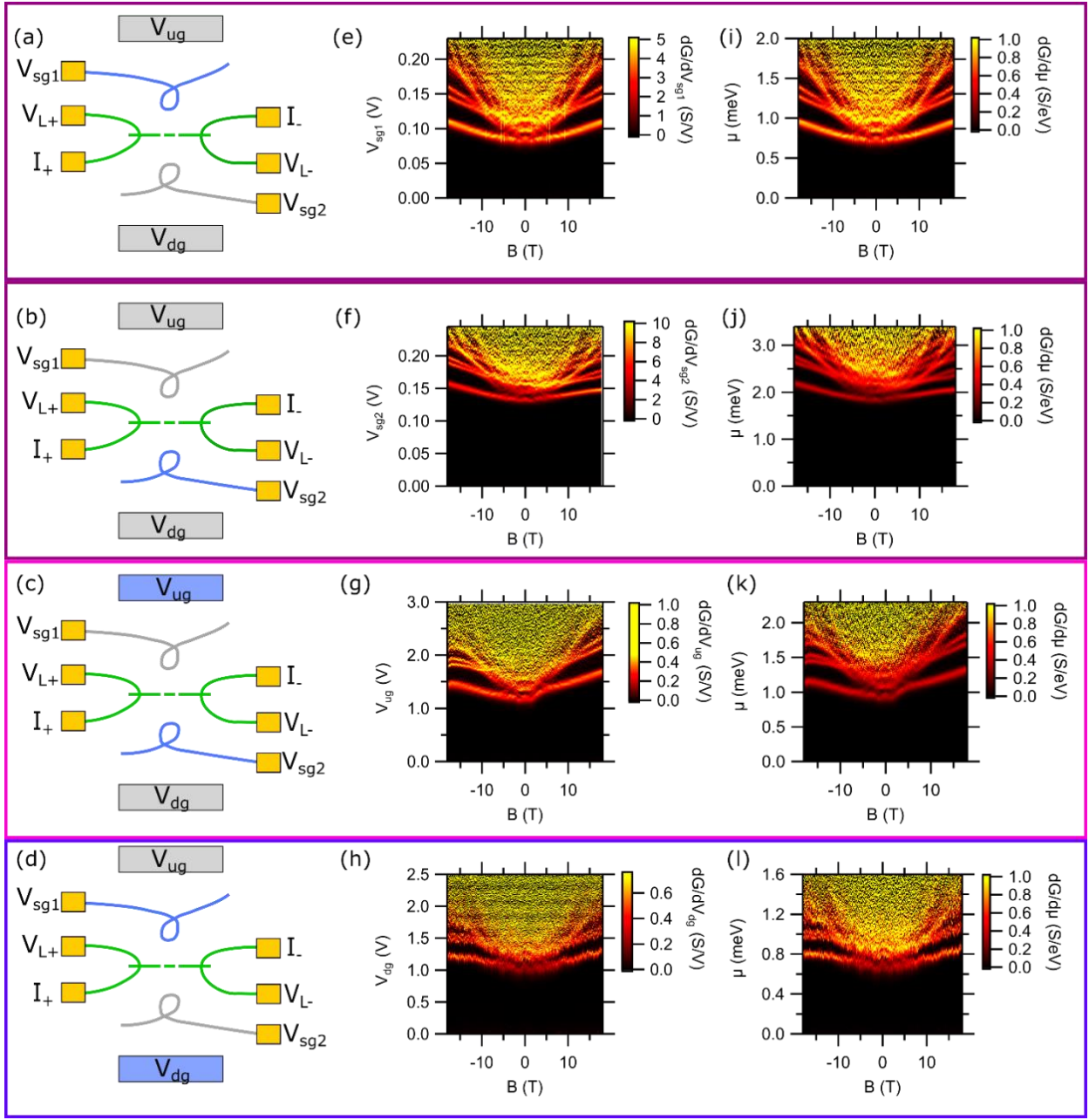

Figure 5: Transconductance measurements across device A for the four gate configurations (a-d) Green region shows the main device, blue region shows the gate voltages applied and grayed regions denotes that the gate is floating. In cases when the up and down gate voltages are being swept, the proximal side gate is maintained at a constant voltage for the whole sweep, (e-h) Transconductance spectra $dG/dV_{xg}$ shown as a function of $V_{xg}$ and $B$ for the four gate configurations, respectively, (i-l) Transconductance spectra $dG/d\mu$ shown as a function of $\mu$ and $B$ for the four gate configurations respectively. All configurations show similar transconductance spectra, irrespective of the distance of the gate from the electron waveguide which ranges from $L_s \sim 2-3$ μm to $L_d \sim 23$ μm. While the side gate tunability range for the up and down gates in 10 times more the proximal side gates, the chemical potential domain is identical for all four cases.

Supplementary Information for:

**Long-Range Nanoelectromechanical Gating at the LaAlO$_3$/SrTiO$_3$ Interface**

Aditi Nethwewala[1,2], Kitae Eom[3], Muqing Yu[1,2], Ranjani Ramachandran[1,2], Chang-Beom Eom[3], Patrick Irvin[1,2], Jeremy Levy[1,2]*

[1]Department of Physics and Astronomy, University of Pittsburgh, Pittsburgh, PA 15260, USA.

[2]Pittsburgh Quantum Institute, Pittsburgh, PA, 15260 USA.

[3]Department of Materials Science and Engineering, University of Wisconsin-Madison, Madison, WI 53706, USA.

*To whom correspondence should be addressed. E-mail: jlevy@pitt.edu

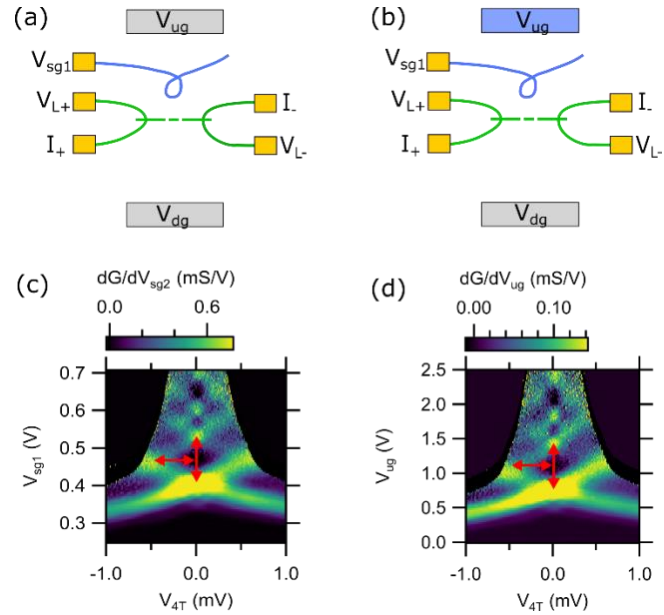

Figure S1: Finite Bias Spectroscopy on Device B (a, b) Schematic showing the device geometry and gate voltages applied. Green region shows the main device, blue region shows the gate voltages applied and grayed regions denotes that the gate is floating. In cases when the up-gate voltage is being swept, the proximal side gate is maintained at a constant voltage for the whole sweep, (c, d) IV measurements across the different gate configuration. Red arrows denote the parameters used to calculate the lever arm for each configuration.

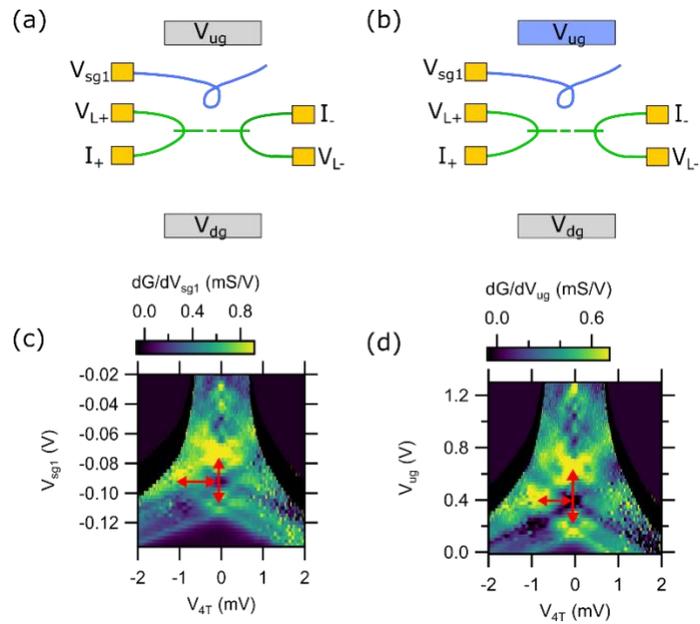

Figure S2: Finite Bias Spectroscopy on Device C (a, b) Schematic showing the device geometry and gate voltages applied. Green region shows the main device, blue region shows the gate voltages applied and grayed regions denotes that the gate is floating. In cases when the up gate voltages is being swept, the proximal side gate is maintained at a constant voltage for the whole sweep, (c, d) IV measurements across the different gate configuration. Red arrows denote the parameters used to calculate the lever arm for each configuration.

# Ratio of Lever Arms

Lever arm ratio is defined as

$$\alpha_x = \frac{C_{xg}}{C_\Sigma},$$

where, $C_{xg}$ is the geometrical gate capacitance and $C_\Sigma$ is the self-capacitance. Self-capacitance is constant for a given device (the electron waveguides here) and the geometrical capacitance vary with the device geometry and spatial distances. The geometrical capacitance and lever arm ratios for Device A is calculated based on the assumption that the host material is a semiconductor. In case of a semiconducting material, geometrical capacitance of a side gate is given by:

$$C_{sg1} = \Sigma \frac{\varepsilon}{\pi} \frac{A_\omega A_{sg1}}{r_{sg1}^3}$$

where, $\varepsilon$ is the dielectric constant, $r_{sg1}$ is the distance between the side gate components and the waveguide, $A_{sg1}$ is the area of the side gate components, and $A_\omega$ is the area of the waveguide. We perform a geometrical sum to account for the components shown in the Figure S3.

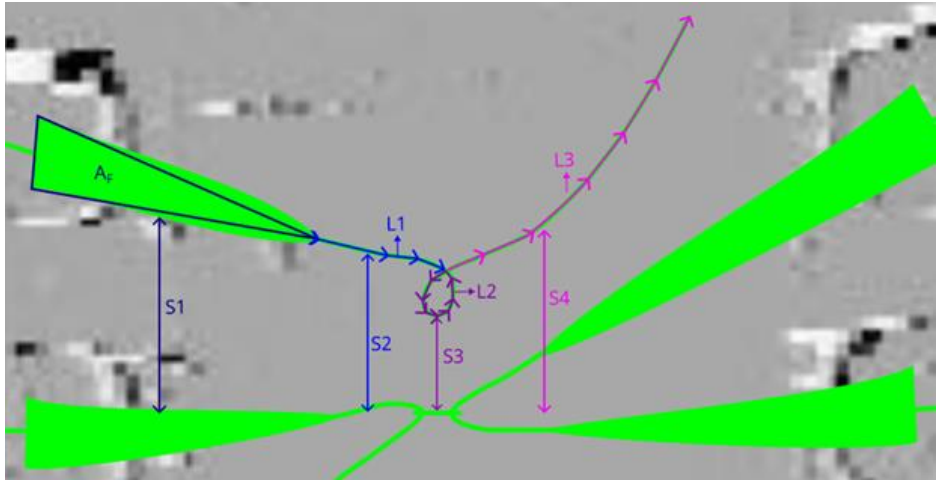

Figure S3: Different components of proximal side gate voltage, $V_{sg1}$, considered while calculating the geometrical capacitance $C_{sg1}$.

Table 1: Area and length of the different components of proximal side gate voltage, $V_{sg1}$, considered while calculating the geometrical capacitance $C_{sg1}$ as highlighted in Figure S3.

| Symbol | Description | Value |
|---|---|---|
| $A_F$ | Approximate area of funnel | 4.6 µm² |
| S1 | Distance between segment 1 and waveguide | 3.96 µm |
| S2 | Distance between segment 2 and waveguide | 3.25 µm |
| S3 | Distance between segment 3 and waveguide | 1.88 µm |
| S4 | Distance between segment 4 and waveguide | 3.78 µm |
| L1 | Length of segment 1 | 2.72 µm |
| L2 | Length of segment 2 | 2.39 µm |
| L3 | Length of segment 3 | 7.42 µm |

Hence,

$$C_{sg1} = 0.0745 \frac{\varepsilon A_\omega}{\pi}.$$

and the lever arm ratio for the side gate is given by:

$$\alpha_{sg1} = \frac{\varepsilon A_\omega}{\pi C_\Sigma} 0.0745.$$

Similarly, lever arm ratio for the up gate can be calculated as:

$$\alpha_{ug} = \frac{\varepsilon A_\omega}{\pi C_\Sigma} 0.0169.$$

We calculate the ratio of $\alpha_{sg1}$ and $\alpha_{ug}$ to obtain a purely geometrical quantity, given by

$$\frac{\alpha_{sg1}}{\alpha_{ug}} = 4.4.$$

The above ratio obtained for a semiconducting host material is supposed to be a constant for all devices since it only depends on the device geometry. However, as shown in Figure 3, this ratio varies between 6 – 21 for the ballistic electron waveguides written at the LAO/STO interface.